\documentclass[onecolumn,amsmath,showpacs,nofootinbib,11pt]{revtex4-2}
\usepackage{graphicx}
\usepackage{dcolumn}
\usepackage{bm}
\begin{document}
\newcommand{\hs}{\hspace*{0.5cm}}
\newcommand{\vs}{\vspace*{0.5cm}}
\newcommand{\be}{\begin{equation}}
\newcommand{\ee}{\end{equation}}
\newcommand{\bea}{\begin{eqnarray}}
\newcommand{\eea}{\end{eqnarray}}
\newcommand{\ben}{\begin{enumerate}}
\newcommand{\een}{\end{enumerate}}
\newcommand{\bde}{\begin{widetext}}
\newcommand{\ede}{\end{widetext}}
\newcommand{\nn}{\nonumber}
\newcommand{\crn}{\nonumber \\}
\newcommand{\Tr}{\mathrm{Tr}}
\newcommand{\non}{\nonumber}
\newcommand{\noi}{\noindent}
\newcommand{\al}{\alpha}
\newcommand{\la}{\lambda}
\newcommand{\bet}{\beta}
\newcommand{\ga}{\gamma}
\newcommand{\va}{\varphi}
\newcommand{\om}{\omega}
\newcommand{\pa}{\partial}
\newcommand{\+}{\dagger}
\newcommand{\fr}{\frac}
\newcommand{\bc}{\begin{center}}
\newcommand{\ec}{\end{center}}
\newcommand{\Ga}{\Gamma}
\newcommand{\de}{\delta}
\newcommand{\De}{\Delta}
\newcommand{\ep}{\epsilon}
\newcommand{\varep}{\varepsilon}
\newcommand{\ka}{\kappa}
\newcommand{\La}{\Lambda}
\newcommand{\si}{\sigma}
\newcommand{\Si}{\Sigma}
\newcommand{\ta}{\tau}
\newcommand{\up}{\upsilon}
\newcommand{\Up}{\Upsilon}
\newcommand{\ze}{\zeta}
\newcommand{\ps}{\psi}
\newcommand{\Ps}{\Psi}
\newcommand{\ph}{\phi}
\newcommand{\vph}{\varphi}
\newcommand{\Ph}{\Phi}
\newcommand{\Om}{\Omega}
\newcommand{\AdrHEPC}{Phenikaa Institute for Advanced Study and Faculty of Basic Science, Phenikaa University, Yen Nghia, Ha Dong, Hanoi 100000, Vietnam}

\title{Novel imprint of a dark photon from the 3-3-1-1 model}

\author{Doan Minh Luong}
\email{21010345@st.phenikaa-uni.edu.vn}
\affiliation{Phenikaa Institute for Advanced Study, Phenikaa University, Yen Nghia, Ha Dong, Hanoi 100000, Vietnam}
\author{Phung Van Dong}
\email{dong.phungvan@phenikaa-uni.edu.vn (corresponding author)}
\affiliation{Phenikaa Institute for Advanced Study, Phenikaa University, Yen Nghia, Ha Dong, Hanoi 100000, Vietnam}
\author{Nguyen Huy Thao}
\email{nguyenhuythao@hpu2.edu.vn}
\affiliation{Department of Physics, Hanoi Pedagogical University 2, Phuc Yen, Vinh Phuc, Vietnam}

\date{\today}

\begin{abstract}
We investigate a dark photon that arises from the UV model based upon $SU(3)_C\otimes SU(3)_L\otimes U(1)_X \otimes U(1)_G$ (3-3-1-1) gauge symmetry, where
the last three factors enlarge the electroweak symmetry encompassing  
electric charge $Q=T_3 - 1/ \sqrt{3}T_8 +X$ and dark charge $D = -2/\sqrt{3} T_8 +G$. It is well-established that this model addresses the questions of family number, neutrino mass, and dark matter. It is shown in this work that if the 3-3-1-1 breaking scale is much bigger than the dark charge breaking scale, the relevant dark gauge boson $Z'$ is uniquely imprinted at TeV, avoiding dangerous FCNC processes, obeying precision electroweak measurements, as well as contributing to collider phenomena, even if no kinetic mixing is presented. The dark matter observables are perhaps governed by the dark charge breaking Higgs field instead of the dark photon. 
\end{abstract}

\maketitle

\section{Introduction}

The model based upon $SU(3)_C\otimes SU(3)_L\otimes U(1)_X$ (3-3-1) gauge symmetry is well-motivated as predicting the family number to match that of colors by anomaly cancellation \cite{331v1,331v2,331pp,331f,331flt}. The relevant 3-3-1 symmetry has a special property as combining dark matter and normal matter in gauge multiplets as well as coupling dark matter in pairs in minimal interactions including the gauge interaction. In this model, the $B-L$ charge neither commutes nor closes algebraically with gauge symmetry, which enlarges the current gauge group to a complete symmetry $SU(3)_C\otimes SU(3)_L\otimes U(1)_X\otimes U(1)_N$ (3-3-1-1), where $N$-charge is related to $B-L$ via a $SU(3)_L$ charge \cite{3311v1,3311v2}. The setup of 3-3-1-1 symmetry provides a matter parity as residual gauge symmetry stabilizing a dark matter candidate, since dark matter carries a wrong $B-L$ number and is odd under the matter parity, opposite to normal matter. This new model supplies neutrino mass via seesaw mechanism \cite{3311dm1, 3311dm2, 3311dm3, 3311dm4} or scotogenic mechanism \cite{3311n1,3311n2,3311n3,3311n4,3311n5}. Additionally, the cosmic inflation, baryon asymmetry, and kinetic mixing effect are extensively investigated in \cite{3311km1,3311km2,3311inf1,3311inf2,3311inf3,3311inf4}. Since both dark matter and normal matter carry a $B-L$ charge significantly governed by known weak coupling, the dark matter observables are strictly constrained by experimental detection \cite{3311dm1, 3311dm2, 3311dm3, 3311dm4}. Another issue is that this kind of model supplies dangerous FCNCs induced by both new neutral gauge bosons $Z',Z''$ \cite{3311v2,3311dm1} confronting the current collider searches (see, for instance, \cite{331fcnc1,331fcnc2}).  

Alternative to $B-L$, the recent approach \cite{VanDong:2023lbn} proposes a dark charge $D$ for dark matter, whereas normal matter carries no dark charge. In this case, the dark charge neither commutes nor closes algebraically with gauge symmetry enlarging it to $SU(3)_C\otimes SU(3)_L\otimes U(1)_X\otimes U(1)_G$ (called 3-3-1-1 too) for which $D=-2/\sqrt{3} T_8+G$. This interpretation of dark charge yields naturally a scotogenic scheme for neutrino mass and dark matter, instead of the canonical seesaw in $B-L$ setup. It is noteworthy that this theory unifies a dark photon coupled to $D$ \cite{phdp} with electroweak gauge bosons in the same manner the electroweak theory \cite{ewt1,ewt2,ewt3} set for photon and weak gauge bosons, called scotoelectroweak unification. Interestingly, if the scale of scotoelectroweak unification is high, much bigger than the dark charge scale at TeV, the 3-3-1-1 symmetry is reduced to $SU(3)_C\otimes SU(2)_L\otimes U(1)_Y\otimes U(1)_D$, where $D$ as given is family universal, similarly to $Y=-1/\sqrt{3} T_8+X$. This suppresses the large FCNCs exchanged by new neutral gauge bosons, as well-established in 3-3-1 and 3-3-1-1 models, even if the dark photon associated with $D$ is at TeV. Additionally, this unification of dark charge predicts precisely dark photon couplings with normal fermions appropriate to the precision data, thus predictive signals of dark photon at current colliders. 

The rest of this work is organized as follows. In Sec. \ref{scotoelectroweak}, we reconsider the scotoelectroweak model. In Sec. \ref{darkphoton}, we obtain novel imprint of the dark photon at TeV. In Sec. \ref{fcnc}, we investigate the precision electroweak measurements affected by dark photon. In Sec. \ref{collider}, we estimate dark photon signals at LEPII and LHC. In Sec. \ref{nudar}, we revisit the issues of neutrino mass and dark matter. Finally, we conclude this work in Section \ref{conclusion}.

\section{\label{scotoelectroweak} The scotoelectroweak unification}

The standard model arranges left-handed fermions in doublets $(\nu_{L},e_{L})\sim 2$ and $(u_{L},d_{L})\sim 2$ while it puts right-handed fermions in singlets $e_{R}\sim 1$, $u_{R}\sim 1$, and $d_{R}\sim 1$. Extending $SU(2)_L$ to $SU(3)_L$ correspondingly enlarges fermion doublets to a fundamental triplet (3) or antitriplet ($3^*$), whereas fermion singlets are retained. The $[SU(3)_L]^3$ anomaly cancellation requires the number of triplets to equal that of antitriplets. This implies the family number to be a multiple of color number,~3. Hence, the 3 families uniquely exist ensuring QCD asymptotic freedom that limits the family number to be less than or equal to 5.   

The $SU(3)_L$ multiplets decompose as $3=2\oplus 1$ and $3^*=2^*\oplus 1$ under $SU(2)_L$. Enlarging usual fermion doublets ($2/2^*$) implies new fermion singlets (1's) being at bottom of $3/3^*$, such as   
\bea \psi_{aL} = \begin{pmatrix}\nu_{aL}\\
e_{aL}\\
N_{aL}
\end{pmatrix}\sim 3,\hs
Q_{\al L} = \begin{pmatrix}d_{\al L}\\
-u_{\al L}\\
D_{\al L}
\end{pmatrix}\sim 3^*,\hs
Q_{3 L} = \begin{pmatrix} u_{3L}\\
d_{3L}\\
U_{3L}
\end{pmatrix}\sim 3,\eea where $a=1,2,3$ and $\al=1,2$ are family indices. Right-handed fermions are $SU(3)_L$ singlets, 
\be e_{aR}\sim 1,\hs N_{aR}\sim 1,\hs u_{aR}\sim 1,\hs d_{aR}\sim 1,\hs D_{\al R}\sim 1,\hs U_{3R}\sim 1. \ee 

We assume that $N_a$, $D_\al$, and $U_3$ possess a dark charge $D=1$, $-1$, and 1, respectively, whereas all the normal fields have no dark charge, i.e. $D=0$, as all collected in Tab.~\ref{tab1}~\cite{VanDong:2023lbn}. Additionally, we assume that $N_a$, $D_\al$, and $U_3$ have an electric charge $Q=0$, $-1/3$, and $2/3$, respectively, like those of the 3-3-1 model with right-handed neutrinos.

\begin{table}[h]
\bc
\begin{tabular}{lccccccccccccccccccccccc}
\\ \hline\hline 
Particle & $\nu_a$ & $e_a$ & $N_a$ & $u_a$ & $d_a$ & $D_\al$ & $U_3$ & $\chi$ & $\eta_{1,2}$& $\rho_{1,2}$ & $\varphi_3$ & $\eta_3$ & $\rho_3$ & $\varphi_{1,2}$ & $\phi$ & gluon & $\ga$ & $Z$ & $Z'$ & $Z''$ & $W$ & $X^0$ & $Y^-$ \\
\hline 
$D$ & 0 & 0 & 1 & 0 & 0 & $-1$ & $1$ & 1 & 0 & 0 & 0 & $1$& $1$ & $-1$ & $-2$ & 0 & 0 & 0 & 0 & 0 & 0 & $-1$ & $-1$\\
\hline
$P_D$ & $+$ & $+$ & $-$ & $+$ & $+$ & $-$ &$-$ &$-$ & $+$ &$+$ &$+$ &$-$ &$-$ &$-$ & $+$ &$+$ &$+$ &$+$ &$+$ &$+$ & $+$ & $-$ &$-$ \\
\hline\hline
\end{tabular}
\caption[]{\label{tab1} Dark charge ($D$) and dark parity ($P_D$).}
\ec
\end{table}

We derive $Q=\mathrm{diag}(0,-1,0)$ and $D=\mathrm{diag}(0,0,1)$, for a lepton triplet $\psi_L$, which neither commute nor close algebraically with $SU(3)_L$. New abelian charges $X$ and $G$ arise by symmetry principle, yielding a 3-3-1-1 gauge symmetry,
\be SU(3)_C\otimes SU(3)_L \otimes U(1)_X\otimes U(1)_G, \ee where the color group is also included, and $X,G$ determine $Q,D$, such as 
 \be Q=T_3-\fr{1}{\sqrt{3}}T_8+X,\hs D=-\fr{2}{\sqrt{3}}T_8+G,\label{qd}\ee where $T_{n}$ ($n=1,2,3,\cdots,8$) denotes $SU(3)_L$ charges. 
 
The fermion representations under the 3-3-1-1 symmetry are given by  
\bea && \psi_{aL}\sim (1,3,-1/3,1/3),\hs Q_{\al L}\sim (3,3^*,0,-1/3), \hs Q_{3L}\sim (3,3,1/3,1/3),\\
&& e_{aR}\sim (1,1,-1,0),\hs N_{aR}\sim (1,1,0,1),\hs u_{aR}\sim (3,1,2/3,0),\\ 
&&d_{aR}\sim (3,1,-1/3,0),\hs D_{\al R}\sim (3,1,-1/3,-1),\hs U_{3R}\sim (3,1,2/3,1).\eea It is shown that all other anomalies vanish, as desirable \cite{VanDong:2023lbn}. 

The gauge symmetry breaking is derived by scalar fields,   
\bea \eta &=& \begin{pmatrix} \eta^0_1 \\
\eta^-_2 \\ \eta^0_3 \end{pmatrix} \sim (1,3,-1/3,1/3),\hs 
\rho = \begin{pmatrix} \rho^+_1 \\
\rho^0_2 \\ \rho^+_3 \end{pmatrix} \sim (1,3,2/3,1/3),\\ 
\varphi &=& \begin{pmatrix} \varphi^0_1 \\
\varphi^-_2 \\ \varphi^0_3 \end{pmatrix} \sim (1,3,-1/3,-2/3),\hs
\phi \sim (1,1,0,-2),\eea which possess a VEV, 
\bea \langle \eta\rangle  &=& \begin{pmatrix} \fr{u}{\sqrt{2}} \\
0 \\ 0 \end{pmatrix},\hs  
\langle \rho\rangle  = \begin{pmatrix} 0 \\
\fr{v}{\sqrt{2}} \\ 0 \end{pmatrix},\hs 
\langle \varphi\rangle  = \begin{pmatrix} 0 \\
0 \\ \fr{w}{\sqrt{2}} \end{pmatrix}, \hs
\langle \phi\rangle  = \fr{\La}{\sqrt{2}}.\eea The dark charge values of scalar components are shown in Tab. \ref{tab1}.

The fields $\eta$, $\rho$, and $\varphi$ break $SU(3)_L\otimes U(1)_X\otimes U(1)_G$ down to $U(1)_Q\otimes U(1)_D$, while the field $\phi$ breaks $U(1)_D$ down to a dark parity $P_D$ \cite{VanDong:2023lbn}, \be P_D=(-1)^{D}=(-1)^{-\fr{2}{\sqrt{3}}T_8+G}.\label{dp}\ee Notice that $\eta^0_3$ and $\varphi^0_1$ do not develop a VEV as prevented by dark parity conservation. A scalar field $\xi \sim (1,1,0,1)$ that is odd under $P_D$ was included in \cite{VanDong:2023lbn} for scotogenic neutrino mass generation, but it does not affect the current consideration and result and is thus skipped. Given that $w\gg \La \gg u,v$, the scheme of symmetry breaking is
\bc
\begin{tabular}{c}
$SU(3)_C \otimes SU(3)_L \otimes U(1)_X \otimes U(1)_G$\\
$\downarrow w $\\
$SU(3)_C \otimes SU(2)_L \otimes U(1)_Y \otimes U(1)_D$\\
$\downarrow \La $\\
$SU(3)_C \otimes SU(2)_L \otimes U(1)_Y \otimes P_D$\\
$\downarrow u,v $\\
$SU(3)_C \otimes U(1)_Q \otimes P_D$\\
\end{tabular} \ec Here, when $w\gg \La$, the 3-3-1-1 model presents a UV-completion of the often-studied dark charge, while $\La\gg u,v$ ensures a consistency with the standard model. 

The total Lagrangian has the form, $\mathcal{L}=\mathcal{L}_{\mathrm{kin}}+\mathcal{L}_{\mathrm{Yuk}}-V$. The kinetic term is   
\bea \mathcal{L}_{\mathrm{kin}}&=&\sum_F\bar{F}i\ga^\mu D_\mu F + \sum_S(D^\mu S)^\dagger (D_\mu S) -\fr{1}{4}\sum_A A_{\mu\nu}A^{\mu\nu},\eea where $F,S,A$ sum over fermion, scalar, gauge multiplets, respectively, for which the covariant derivative $D_\mu$ is defined by \be D_\mu = \pa_\mu + ig_s t_n G_{n\mu} + ig T_n A_{n\mu} + ig_X X B_\mu + i g_G G C_\mu,\ee  where ($g_s,\ g,\ g_X,\ g_G$), ($G_{n\mu}, A_{n\mu}, B_\mu, C_\mu$), and ($t_n,\ T_n,\ X,\ G$) are couplings, gauge bosons, and charges according to 3-3-1-1 groups, respectively. 

The Yukawa Lagrangian is  
\bea \mathcal{L}_{\mathrm{Yuk}}&=&h^e_{ab}\bar{\psi}_{aL}\rho e_{bR} +h^N_{ab}\bar{\psi}_{aL}\varphi N_{bR}+\fr 1 2 h'^N_{ab}\bar{N}^c_{aR} N_{bR}\phi \crn
&& + h^d_{\al a} \bar{Q}_{\al L}\eta^* d_{aR} +h^u_{\al a } \bar{Q}_{\al L}\rho^* u_{aR} + h^D_{\al \beta}\bar{Q}_{\al L} \varphi^* D_{\beta R}\crn
&&+ h^u_{3a} \bar{Q}_{3L}\eta u_{aR}+h^d_{3a}\bar{Q}_{3L}\rho d_{aR}+ h^U_{33}\bar{Q}_{3L}\varphi U_{3R} +H.c.,\eea while the scalar potential is 
\bea
V &=& \mu^2_1\rho^\dagger \rho + \mu^2_2 \varphi^\dagger \varphi + \mu^2_3 \eta^\dagger \eta + \la_1 (\rho^\dagger \rho)^2 + \la_2 (\varphi^\dagger \varphi)^2 + \la_3 (\eta^\dagger \eta)^2\crn
&&+ \la_4 (\rho^\dagger \rho)(\varphi^\dagger \varphi) +\la_5 (\rho^\dagger \rho)(\eta^\dagger \eta)+\la_6 (\varphi^\dagger \varphi)(\eta^\dagger \eta)\crn
&& +\la_7 (\rho^\dagger \varphi)(\varphi^\dagger \rho) +\la_8 (\rho^\dagger \eta)(\eta^\dagger \rho)+\la_9 (\varphi^\dagger \eta)(\eta^\dagger \varphi) + (f\epsilon^{ijk}\eta_i\rho_j\varphi_k+H.c.) \crn
&& + \mu^2 \phi^\dagger \phi + \la (\phi^\dagger \phi)^2 +\la_{10} (\phi^\dagger \phi)(\rho^\dagger\rho)+\la_{11}(\phi^\dagger \phi)(\varphi^\dagger \varphi)+\la_{12}(\phi^\dagger \phi)(\eta^\dagger \eta).\eea Above, $h$'s and $\la$'s are dimensionless couplings, whereas $\mu$'s and $f$ are mass parameters. 

\section{\label{darkphoton} Novel imprint of the dark photon}

The new leptons $N_{1,2,3}$ gain a pseudo-Dirac mass at $w$ scale due to $w\gg \La$, while the new quarks $D_{1,2}$ and $U_3$ get a Dirac mass also at $w$ scale. They are all integrated out. Concerning the scalar sector, given that $f\sim w\gg \La \gg u,v$, all new Higgs fields receive a large mass at $w$ scale being thus integrated out, except a new neutral Higgs boson $H' \sim \Re(\phi)$ associated with dark charge breaking has a mass at $\La$ scale, the standard model Higgs boson $H \sim u\Re(\eta^0_1)+v\Re(\rho^0_2)$ with a mass at weak scale, and massless Goldstone bosons $G_{Z'}\sim \Im(\phi)$, $G_Z\sim u\Im(\eta^0_1)-v\Im(\rho^0_2)$, and $G^\pm_W\sim u \eta^\pm_2 - v \rho^\pm_1$ associated with $Z'$, $Z$, and $W^\pm$, respectively (cf. \cite{3311dm1}).    

Similarly, gauge bosons obtain a mass from $\mathcal{L}\supset \sum_S(D^\mu \langle S\rangle)^\dagger (D_\mu \langle S\rangle)$ for which substituting the VEVs, we get physical non-Hermitian fields
\be W^\pm = \fr{A_{1}\mp i A_{2}}{\sqrt{2}},\hs X^{0,0*}=\fr{A_{4}\mp i A_{5}}{\sqrt{2}},\hs Y^\mp = \fr{A_{6}\mp i A_{7}}{\sqrt{2}},\ee with respective masses,
\be m^2_W=\fr{g^2}{4}(u^2+v^2),\hs m^2_{X}=\fr{g^2}{4}(u^2+w^2),\hs m^2_Y=\fr{g^2}{4}(v^2+w^2).\ee Here, $W$ is identical to that of the standard model and $u^2+v^2=(246\ \mathrm{GeV})^2$, whereas $X,Y$ are heavy at $w$ scale and are thus integrated out. 

Concerning neutral gauge bosons, we first identify the photon field that is coupled to the electric charge $Q=T_3-1/\sqrt{3}T_8+X$ to be \be A = s_W A_{3}+c_W\left(-\fr{t_W}{\sqrt{3}}A_{8}+\sqrt{1-\fr{t^2_W}{3}}B \right), \ee where the sine of the Weinberg angle is matched by $s_W=e/g=\sqrt{3} t_X/\sqrt{3+4t^2_X}$ with $t_X=g_X/g$, and notice that the field in parentheses is coupled to the hypercharge $Y=-1/\sqrt{3}T_8+X$ \cite{dl}. We next identify the standard model $Z$ boson,
\be Z = c_W A_{3}-s_W\left(-\fr{t_W}{\sqrt{3}}A_{8}+\sqrt{1-\fr{t^2_W}{3}}B \right),\ee which is orthogonal to $A$, as usual, while the 3-3-1 model neutral gauge boson,  
\be \mathcal{Z}' = \sqrt{1-\fr{t^2_W}{3}}A_{8}+\fr{t_W}{\sqrt{3}}B,\ee is given/orthogonal to the hypercharge field. 

The photon $A$ is massless and decoupled as a physical field. By contrast, $Z$ and $\mathcal{Z}'$ mix with $C$-boson of $U(1)_G$ via a mass matrix,   
\bea && M^2 =
\left(
\begin{array}{ccc}
m^2_{Z} & m^2_{Z\mathcal{Z}'} & m^2_{ZC}\\
m^2_{Z\mathcal{Z}'} & m^2_{\mathcal{Z}'} & m^2_{\mathcal{Z}'C}\\
m^2_{ZC} & m^2_{\mathcal{Z}'C} & m^2_{C}
\end{array}
\right)= \crn
&&
\fr{g^2}{2} \left(
  \begin{array}{ccc}
    \fr {(3 + 4 t_X^2) (u^2 + v^2)}{ 2 (3 + t_X^2)}
    &  \fr{\sqrt{3 + 4 t_X^2} ((3 - 2 t_X^2) u^2 - (3 + 4 t_X^2) v^2)}
    {6 (3 + t_X^2)}
    & \fr {\sqrt{3 + 4 t_X^2} t_G (u^2-v^2)}{3 \sqrt{3 + t_X^2}} \\
    \fr{\sqrt{3 + 4 t_X^2} ((3 - 2 t_X^2) u^2 - (3 + 4 t_X^2) v^2)}
    {6 (3 + t_X^2)}
    & \fr {(3 -2 t_X^2)^2 u^2 + (3 + 4 t_X^2)^2 v^2 + 4(3 + t_X^2)^2 w^2}{18 (3 + t_X^2)}
    & \fr {t_G ((3-2 t_X^2) u^2 + (3 + 4 t_X^2) v^2 + 4 (3 + t_X^2) w^2)}{9\sqrt{3 + t_X^2}} \\
    \fr {\sqrt{3 + 4 t_X^2} t_G (u^2-v^2)}{3 \sqrt{3 + t_X^2}}
    & \fr {t_G ((3-2 t_X^2) u^2 + (3 + 4 t_X^2) v^2 + 4 (3 + t_X^2) w^2)}{9\sqrt{3 + t_X^2}}
    & \fr 2 9 t_G^2 (u^2 + v^2 + 4 (w^2 +9 \La^2)) \\
  \end{array}
\right),\nn\eea given in the $(Z,\mathcal{Z}',C)$ basis, where $t_G=g_G/g$. Because of $u^2,v^2\ll \La^2\ll w^2$, the field $Z$ is light and decoupled from $\mathcal{Z}',C$ to be a physical field with mass, 
\be m^2_Z\simeq \fr{g^2}{4c^2_W}(u^2+v^2),\ee whereas $\mathcal{Z}',C$ significantly mix, yielding physical fields,
\be Z'=c_\theta \mathcal{Z}'-s_\theta C,\hs Z''=s_\theta \mathcal{Z}'+c_\theta C,\ee with mixing angle and respective masses,
\be t_{\theta}\simeq \fr{\sqrt{3+t^2_X}}{2t_G},\hs m^2_{Z'}\simeq \fr{4g^2_G(3+t_X^2)}{4t^2_G+3+t^2_X}\La^2,\hs m^2_{Z''}\simeq \fr{g^2}{9}(4 t^2_G+3+t^2_X)w^2. \label{zpzppmixmass}\ee The $Z''$ boson is associated with 3-3-1-1 symmetry, heavy at $w$ scale and integrated out. The $Z'$ boson is coupled to the dark charge $D=-2/\sqrt{3}T_8+G$, having a mass at $\La\sim$ TeV scale.
 
In summary, the new physics at TeV scale is only governed by $U(1)_D$ with its gauge field $Z'$ and Higgs field $H'$, while all other new particles ($N_a$, $U_3$, $D_\al$, $X$, $Y$, $Z''$, and the rest of new Higgs fields) are superheavy as manifestly integrated out. The relevant interactions of $Z'$ with normal fermions are straightforwardly computed as 
\be \mathcal{L}\supset -\fr 2 3 g_G s_\theta t^2_W \sum_f \bar{f}\ga^\mu Y f Z'_\mu, \label{effzp}\ee where $f$ runs over usual fermion doublets and singlets. These couplings are proportional to the hypercharge $Y$, resulting from a mixing between $\mathcal{Z}'$ and $C$ due to symmetry breaking, not a kinetic mixing as in the normal sense. For convenience, the couplings are collected in Tab. \ref{tab2} too.
\begin{table}[h]
\bc
\begin{tabular}{lcc}
\hline\hline
$f$ & $g^{Z'}_V(f)$ & $g^{Z'}_A(f)$ \\
\hline  
$\nu_a $ & $\fr{-c_\theta s^2_{W}}{2\sqrt{3-4s^2_W}}$  & $\fr{-c_\theta s^2_{W}}{2\sqrt{3-4s^2_W}} $ \\
$e_a$ & $\fr{-3c_\theta s^2_W }{2\sqrt{3-4s^2_W}} $ & $\fr{c_\theta s^2_W}{2\sqrt{3-4s^2_W}} $ \\  
$u_a$ & $\fr{5c_\theta s^2_W}{6\sqrt{3-4s^2_W}}$ & $\fr{-c_\theta s^2_W}{2\sqrt{3-4s^2_W}} $ \\ 
$d_a$ & $\fr{-c_\theta s^2_W}{6\sqrt{3-4s^2_W}}$ & $\fr{c_\theta s^2_{W}}{2\sqrt{3-4s^2_W}} $\\ 
\hline\hline
\end{tabular}
\caption{\label{tab2} Couplings of $Z'$ dark photon with fermions.}
\ec
\end{table}  A final remark is that the dark photon does not flavor change, although it originates from a family-dependent gauge charge $T_8$. This avoids the neutral meson mixing constraint, opposite to those in 3-3-1 and 3-3-1-1 models \cite{3311v2,3311dm1,331fcnc1,331fcnc2}. 

\section{\label{fcnc} EWPT affected by dark photon}

The precision data consist of low-energy weak neutral current, $Z$-pole physics, and others. The weak neutral current is affected by $Z'$ exchange, mainly
sensitive to $Z'$ mass. The $Z$-pole physics measured at LEP and SLC is mainly sensitive to $Z$-$(\mathcal{Z}',C)$ mixing, modifying $Z$ mass and $Zf\bar{f}$ couplings, by contrast. The precision data strongly limit $Z$-$(\mathcal{Z}',C)$ mixing angles, say $\mathcal{E}\sim 10^{-3}$, extracted by comparing coupling measurements with standard model prediction (cf. \cite{zzprimemixing}). This also limits new neutral gauge boson mass but generally weaker than high energy collider, e.g. LEPII and LHC, which is above $Z$ pole and insensitive to mixing angles $\mathcal{E}$. Alternatively, global fit to full electroweak data set yields a comparable bound as $\mathcal{E}\sim 10^{-3}$ \cite{lepewwg}.  

\subsection{Atomic parity violation}

It is clear from (\ref{effzp}) that the dark photon couplings to usual fermions violate parity at a considerable level because the left-handed and right-handed fermions including neutrinos have different hypercharges. In the conventional dark photon model, the couplings of dark photon with fermions are vectorlike, conserving parity, by contrast. This is a new observation of the work, allowing discriminating the current model in experiment. 

Unlike the conventional dark photon, the current field $Z'$ contributes to atomic parity violation via an effective interaction,
\be \mathcal{L}_{\mathrm{eff}}\supset \fr{G_F}{\sqrt{2}}(\bar{e}\ga_\mu \ga_5 e)(C'_{1u} \bar{u}\ga^\mu u + C'_{1d}\bar{d}\ga^\mu d),\ee where 
\be \fr{G_F}{\sqrt{2}}=\fr{1}{2(u^2+v^2)},\hs C'_{1u}= -\fr{5c^2_\theta s^4_W}{6(3-4s^2_W)}\fr{m^2_Z}{m^2_{Z'}},\hs C'_{1d}=\fr{c^2_\theta s^4_W}{6(3-4s^2_W)}\fr{m^2_Z}{m^2_{Z'}}.\ee A parity violation that couples electrons as vector and quarks as axial-vector due to $Z'$ exists too, but it is prevented for heavy atoms due to its dependence on spins other than charges, analogous to the case of $Z$, which would be neglected. 

The weak charge deviation from the standard model computed for an atom that composes $Z$ protons and $N$ neutrons is 
\be \Delta Q_W(Z,N)=-2\left[Z\left(2C'_{1u}+C'_{1d}\right)+N\left(C'_{1u}+2C'_{1d}\right)\right]=\fr{(3Z+N)c^2_\theta s^4_W}{3-4s^2_W}\fr{m^2_{Z}}{m^2_{Z'}}.\ee Taking Cesium with $Z=55$ and $N=78$, the current measurement and the standard model value for Cs weak charge make a bound $\Delta Q_W(\mathrm{Cs})<0.61$ \cite{pdg}, implying $m_{Z'}>291 c_\theta$ GeV, for $s^2_W=0.231$ and $m_Z=91.1876$ GeV. Since $c_\theta\leq 1$, this bound is easily evaded by the following constraint, i.e. $Z'$ negligibly contributes to the parity violation in Cs. 

\subsection{$\rho$-parameter}

The mass matrix of $(Z,\mathcal{Z}',C)$ given above has a seesaw form due to $u^2,v^2\ll \La^2\ll w^2$ for which $Z$ is light, while $\mathcal{Z}',C$ are heavy. With the aid of the seesaw diagonalization, we obtain the mass of the physical light state (called $Z_1$), such as 
\be m^2_{Z_1}\simeq m^2_{Z}-\mathcal{E} \begin{pmatrix} m^2_{Z\mathcal{Z}'}\\
m^2_{ZC}
\end{pmatrix},\hs \mathcal{E}\equiv \begin{pmatrix} m^2_{Z\mathcal{Z}'} &
m^2_{ZC} \end{pmatrix} \begin{pmatrix}  m^2_{\mathcal{Z}'} & m^2_{\mathcal{Z}'C}\\
 m^2_{\mathcal{Z}'C} & m^2_{C} \end{pmatrix}.\label{cale} \ee Then, this mass contributes to the $\rho$-parameter at the tree level, such as
\be \Delta \rho = \rho -1 =\fr{m^2_W}{c^2_W m^2_{Z_1}}-1\simeq \fr{t^4_W(u^2+v^2)}{36 \La^2}.\ee  
It is noted that the tree-level terms of $(u^2,v^2)/w^2$ order are negligible and omitted. 

Further, at the one-loop level, the $\rho$-parameter receives a contribution from $(X,Y)$ gauge vector doublet \cite{looprho1,looprho2,looprho3}. However, this radiative correction is proportional to the doublet mass splitting $(m^2_X-m^2_Y)/m^2_{X,Y}\sim (u^2-v^2)/w^2$, which is infinitesimal and suppressed too. 

From the global fit, the $\rho$-parameter is given by $\rho = 1.00038$, which is 2 $\sigma$ above the standard model prediction $\rho=1$ \cite{pdg}. Hence, we require $\Delta \rho<0.00038$, leading to $\La>631$ GeV. This translates to $m_{Z'}=(3gc_W/\sqrt{3-4s^2_W})c_\theta \La>749c_\theta$ GeV for $g=0.65$, which is stronger than the parity violation bound.     

\subsection{$Zf\bar{f}$ couplings}   

The above seesaw diagonalization determines a mixing between $Z$ and $\mathcal{Z}',C$, such as
\be \begin{pmatrix} Z \\
\mathcal{Z}'\\
C\end{pmatrix} = \begin{pmatrix} 
1 & \mathcal{E}_1 & \mathcal{E}_2 \\
-\mathcal{E}_1 & 1 & 0 \\
-\mathcal{E}_2 & 0 & 1\end{pmatrix} \begin{pmatrix} Z_1\\
Z_2\\
Z_3\end{pmatrix},\ee where $\mathcal{E}\equiv (\mathcal{E}_1,\mathcal{E}_2)$ is defined in (\ref{cale}) and evaluated as
\be \mathcal{E}_1=-\fr{t^2_W\sqrt{3-4s^2_W}(u^2+v^2)}{36c^2_W\La^2},\hs \mathcal{E}_2=\fr{t^2_W(u^2+v^2)}{24c_W t_G \La^2},\ee where notice that $|\mathcal{E}_{1,2}|\ll 1$ due to $u^2,v^2\ll \La^2$. It is clear that $Z_1=Z-\mathcal{E}_1 \mathcal{Z}'-\mathcal{E}_2 C\simeq Z$, $Z_2=\mathcal{E}_1 Z+\mathcal{Z}'\simeq \mathcal{Z}'$, and $Z_3=\mathcal{E}_2 Z+C\simeq C$ approximate the original states due to $|\mathcal{E}|\ll 1$. 

$\mathcal{Z}'$ and $C$ modify the well-measured couplings of $Z$ with fermions by amounts $\mathcal{E}_{1,2}$, respectively. Demanding $|\mathcal{E}_{1,2}|<10^{-3}$ gives $\La>972$ and  $929/\sqrt{t_G}$ GeV, respectively. Taking the former bound translates to $m_{Z'}>1.15c_\theta$ TeV, which is stronger than the $\rho$-parameter bound. 

\section{\label{collider} Dark photon signal at colliders}

The precision data hint to a dark photon mass beyond the electroweak scale. Hence, it appropriately looks for a dark photon signal at high energy colliders. Since $Z'$ couples to leptons and quarks, it may be produced at both LEPII and LHC.  

\subsection{LEPII}

The LEPII experiment studied a new neutral gauge boson $Z'$ that mediates the process $e^+e^-\to f\bar{f}$ for $f=\mu,\tau$. The new physics contribution to the process can be determined by integrating $Z'$ out, yielding an effective interaction
\bea \mathcal{L}_{\mathrm{eff}} &=& \fr{g^2}{c^2_W m^2_{Z'}}[\bar{e}\ga^\mu (a^{Z'}_L(e) P_L+a^{Z'}_R(e)P_R)e][\bar{f}\ga_\mu (a^{Z'}_L(f) P_L+a^{Z'}_R(f)P_R)f]\crn
&=&\fr{g^2[a^{Z'}_L(e)]^2}{c^2_W m^2_{Z'}}(\bar{e}\ga^\mu P_L e)(\bar{f}\ga_\mu P_L f)+ (LR)+(RL)+(RR),\label{lepint}\eea since the charged lepton couplings for $f=\mu,\tau$ equal those of electron, and these chiral couplings $a^{Z'}_{L,R}=(g^{Z'}_{V}\pm g^{Z'}_A)/2$ can directly be extracted from Tab. \ref{tab2}.  

The LEPII searched for such chiral interactions as in (\ref{lepint}), providing respective bounds on the couplings, typically \cite{lepii}
\be \fr{g^2[a^{Z'}_L(e)]^2}{c^2_W m^2_{Z'}} <\fr{1}{(6\ \mathrm{TeV})^2}.\ee This implies $m_{Z'}>(3gs^2_W/c_W\sqrt{3-4s^2_W})c_\theta\ \mathrm{TeV}=356c_\theta$ GeV, which is more smaller than the $\rho$-parameter and $f$-fermion coupling bounds. In other words, the dark photon with mass at TeV easily escapes the LEPII detection. 

\subsection{LHC}

\begin{figure}[h]
\bc
\includegraphics[scale=0.8]{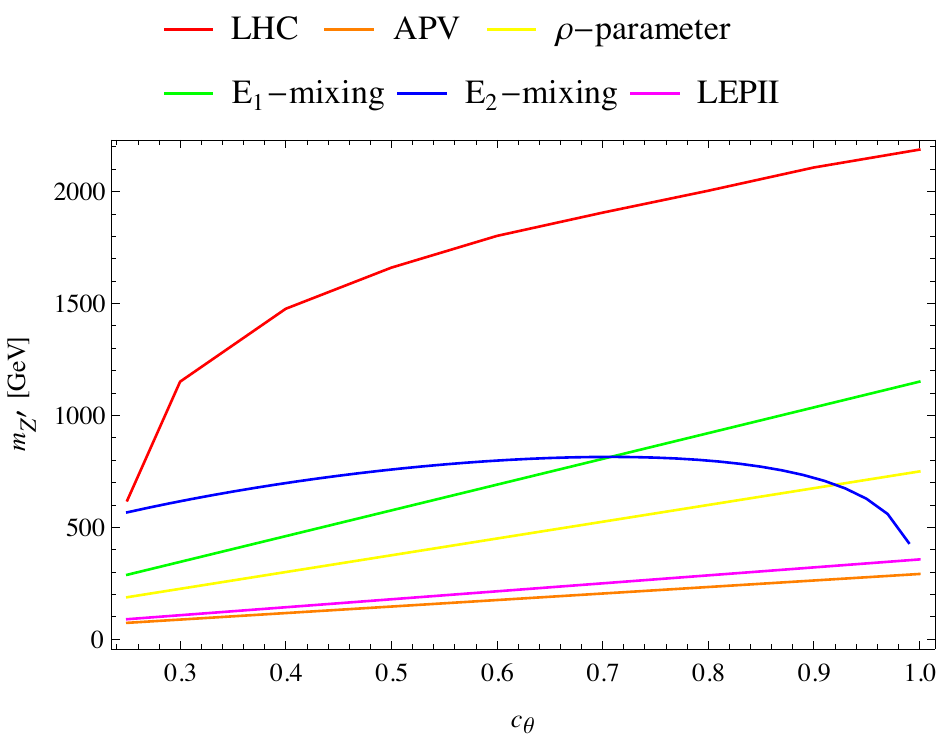}
\caption[]{\label{fig1} The $Z'$ dark photon mass for each value of $c_\theta$ constrained by the LHC dilepton search \cite{lhcdi} (top curve), while the other curves counted from the bottom up present the bounds by atomic parity violation, LEPII, $\rho$-parameter, and $\mathcal{E}_{1,2}$ mixing, respectively.}
\ec
\end{figure}

At the LHC, the dark photon $Z'$ may be produced and then decayed to a dilepton signal, $\sigma(pp \to Z'\to l\bar{l})=\sigma(pp\to Z')\mathrm{Br}(Z'\to l\bar{l})$. This process was particularly studied in \cite{VanDong:2023lbn} for a case with $t_G=1$ or a corresponding $c_\theta$ value following (\ref{zpzppmixmass}) since $t_X=\sqrt{3}s_W/\sqrt{3-4s^2_W}$ is already fixed by the Weinberg angle. We now generalize the result for a variation of $c_\theta$ and obtain the corresponding bound as plotted in Fig. \ref{fig1} (red line), which implies $m_{Z'}>2.18$ TeV for $c_\theta=1$, for a reference. Notice, hereafter, that the allowed regime is given above the relevant curve. 

Further, we also include the atomic parity violation, $\rho$-parameter, $\mathcal{E}_{1,2}$-mixings, and LEPII bounds to Fig. \ref{fig1}, for comparison. For each value of $c_\theta$, the LHC makes the strongest bound on the $Z'$ mass as compared with the other constraints.

\section{\label{nudar} Neutrino mass and dark matter revisited}

Would neutrino mass generation and dark matter stability be determined by the $U(1)_D$ dark physics at TeV regime? Although the issues are still addressed by the model in itself in the case $w\gg \La$ like the current setup as shown in \cite{VanDong:2023lbn}, this work gives an alternative answer. 

The question may be simply answered by introducing three right-handed neutrinos \be \nu_{aR}\sim (1,1,0,0),\ee which both couple to $\psi_{bL}\eta$ and possess a Majorana mass $(m_R)_{ab} \nu_{aR}\nu_{bR}$, inducing the observed neutrino masses via a canonical seesaw, as well as a vectorlike dark fermion \be \chi\sim(1,1,0,1),\ee which provides the lightest component of dark fields ($P_D=-1$) responsible for dark matter. The canonical seesaw works the same normal case \cite{valle}, while the fermion dark matter observables are set by $H'$ portal rather than $Z'$ portal. Let us study the latter, governed by interactions
\be \mathcal{L}\supset \fr{3g c_\theta c_W}{2\sqrt{3-4s^2_W}} \bar{\chi} \ga^\mu \chi Z'_\mu+\left(\mu_\chi \bar{\chi}_L\chi_R +\fr 1 2  h_L \chi_L \chi_L \phi + \fr 1 2 h_R \chi_R \chi_R \phi+H.c.\right), \ee where $\phi=(\La+H'+i G_{Z'})/\sqrt{2}$, as mentioned, while $\mu_\chi$ and $h_{L,R}$ are a mass parameter and dimensionless couplings, respectively. Notice that $\chi$ may insignificantly couple to a superheavy lepton $N$, which is suppressed by a $Z_2$ as $\chi\to-\chi$, without loss of generality. 

Because of $h_{L,R}$ contributions, $\chi$ is generically separated into two physical Majorana fields $\chi_{1,2}$ by diagonalizing, such as  
\bea \mathcal{L} &\supset& -\fr 1 2 \begin{pmatrix} \bar{\chi}_L & \bar{\chi}^c_R\end{pmatrix} \begin{pmatrix} 
-h_L\fr{\La}{\sqrt{2}} & -\mu_\chi \\
-\mu_\chi & -h_R\fr{\La}{\sqrt{2}}\end{pmatrix}\begin{pmatrix} \chi^c_L \\
\chi_R \end{pmatrix}+H.c.\crn
&=& -\fr 1 2 m_{\chi_1} \chi^2_1-\fr 1 2 m_{\chi_2}\chi^2_2+H.c.,\eea where we define $t_{2\zeta} = \fr{2\sqrt{2}\mu_\chi}{(h_R-h_L)\La}$, and \bea &&\chi_1=c_\zeta \chi_L -s_\zeta \chi^c_R,\hs \chi_2= s_\zeta \chi_L +c_\zeta \chi^c_R,\\
&& m_{\chi_{1,2}}=\fr{1}{2\sqrt{2}} \left[-(h_L+h_R)\La\mp \sqrt{(h_L-h_R)^2\La^2+8\mu^2_\chi}\right]. \eea 

We suppose $\chi_1$ to be lightest, i.e. $|m_{\chi_1}|<|m_{\chi_2}|$. Thus, $\chi_1$ is stabilized by $P_D$ to be a dark matter candidate. It possesses the following couplings to $Z',H'$,
\be \mathcal{L}\supset -\fr{3g c_\theta c_W}{4\sqrt{3-4s^2_W}} \bar{\chi}_1 \ga^\mu \ga_5\chi_1 Z'_\mu +\left(\fr 1 2 h_1 \chi_1\chi_1 H'+H.c.\right), \ee where $h_1=(c^2_\zeta h_L+s^2_\zeta h_R)/\sqrt{2}$, and $Z'$ has only axial-vector current to Majorana field $\chi_1$. Due to the Majorana nature, the annihilation of $\chi_1$ via $Z'$ portal is helicity suppressed, which starts only from $p$-wave contributions. Further, $\chi_1$ has only spin-dependent scattering cross-section with nuclei by $Z'$ portal, which is suppressed too (cf. \cite{VanDong:2023lbn}). Thai said, the Higgs portal $H'$ would dominantly contribute to the dark matter observables, as shown below.  

\begin{figure}[h]
\bc
\includegraphics[scale=1]{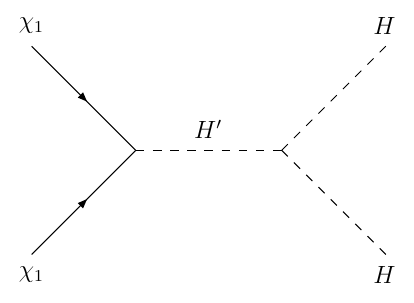}
\caption[]{\label{fig2} Annihilation of Majorana fermion dark matter to usual Higgs fields.}
\ec
\end{figure}     

In the early universe, $\chi_1$ annihilates to the standard model Higgs boson $H$ via the $H'$ portal which sets the relic density, as despited in Fig. \ref{fig2}. It is noted that $H'$ couples to $H$ such as $\mathcal{L}\supset \fr 1 2 \bar{\la}\La H' H^2$, where $\bar{\la}=(\la_{10}v^2+\la_{12}u^2)/(v^2+u^2)$. Therefore, the dark matter annihilation cross-section is straightforwardly computed as
\be \langle \sigma v \rangle_{\chi_1} = \fr{1}{128\pi}\fr{h^2_1 \bar{\la}^2\La^2}{(4m^2_{\chi_1}-m^2_{H'})^2}\sqrt{1-\fr{m^2_H}{m^2_{H'}}}.\ee

\begin{figure}[h]
\bc
\includegraphics[scale=0.9]{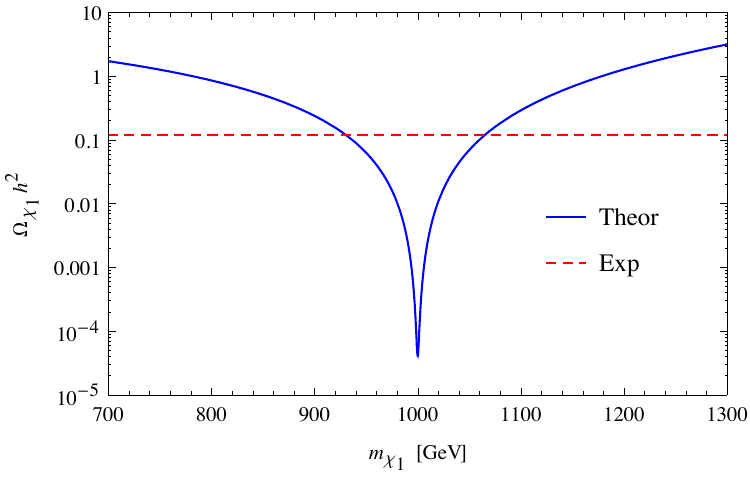}
\caption[]{\label{fig3} Dark matter relic density plotted as function of its mass, where the measured density is also shown.}
\ec
\end{figure}
We take $m_H=125$ GeV, $h_1=0.5$, $\bar{\la}=0.5$, and $\La=2$ TeV, where the last one is appropriate to the strongest bound coming from the LHC. We plot the dark matter relic density $\Om_{\chi_1} h^2\simeq 0.1\ \mathrm{pb}/\langle \sigma v\rangle_{\chi_1}$ as a function of the dark matter mass for a fixed value of the $H'$ mass, say $m_{H'}=2$~TeV, as given in Fig. \ref{fig3}. We also include the experimental value $\Om_{\chi_1} h^2=0.12$ \cite{pdg} to this figure. It is clear from the figure that the $H'$ resonance is crucial to set the relic density. A correct abundance, i.e. $\Om_{\chi_1} h^2\leq 0.12$, requires $m_{\chi_1}=930$--1066 GeV. 

Although $H'$ does not couple to quark, as well as $H$ does not couple to $\chi_1$, it is evident that $\chi_1$ can interact with quark through a $H$-$H'$ mixing given by the scalar potential, since $\La$ is not much bigger than $u,v$. Let us parametrize the mixing, $H_1=c_\delta H - s_\delta H'$ and $H_2=s_\delta H + c_\delta H'$, which $H_1$ now becomes the standard model Higgs like boson, while $H_2$ acts as a new heavy Higgs boson. It is easily derived that $\delta\sim (u,v)/\La$ which arises from the scalar potential. Hence, such mixing gives rise to an effective coupling between $\chi_1$ and quark to be $\mathcal{L}_{\mathrm{eff}}\supset \al^S_q (\bar{\chi}^c_1\chi_{1})(\bar{q}q)$, where 
\be \al^S_q=-\fr{s_\delta c_\delta h_1 m_q}{v_{\mathrm{w}} m^2_{H_1}},\ee which is dominantly contributed by $H_1$, where $v_{\mathrm{w}}=\sqrt{u^2+v^2}=246$ GeV is the weak scale. 

Hence, the scattering cross-section of $\chi_1$ with nucleon $N=p,n$ is  \cite{sis1}
\be \sigma^{\mathrm{SI}}_{\chi_1-N}=\fr{4 m^2_N}{\pi} (f^N)^2,\ee where the form factor is 
\be \fr{f^N}{m_N}=\sum_{q=u,d,s}\fr{\al^S_q}{m_q} f^N_{Tq} +\fr{2}{27} f^N_{TG}\sum_{q=c,b,t} \fr{\al^S_q}{m_q}\simeq -\fr{0.35s_\delta c_\delta h_1}{v_{\mathrm{w}} m^2_{H_1}},\ee where $f^N_{TG}=1-\sum_{q=u,d,s}f^N_{Tq}$, $f^p_{Tu}=0.02$, $f^p_{Td}=0.026$, $f^p_{Ts}=0.118$, $f^n_{Tu}=f^p_{Td}$, $f^n_{Td}=f^p_{Tu}$, and $f^n_{Ts}=f^p_{Ts}$ \cite{sis2}. With the aid of $m_N=1$ GeV and $m_{H_1}=125$ GeV, the SI cross-section becomes 
\be \sigma^{\mathrm{SI}}_{\chi_{1}-N}\simeq \fr{0.49 m^4_N s^2_\delta h^2_1}{\pi v^2_{\mathrm{w}}  m^4_{H_1}}\simeq \left(\fr{s_\varphi}{0.01}\right)^2\left(\fr{h_1}{0.5}\right)^2\times 10^{-46}\ \mathrm{cm}^2.\ee  This cross-section agrees with the latest data $\sigma^{\mathrm{SI}}_{\chi_{1}-N}\sim 10^{-46}\ \mathrm{cm}^2$ \cite{lzexp} for the dark matter mass $m_{\chi_1}\sim$ TeV, the mixing angle $s_\varphi\sim 0.01$, and the coupling constant $h_1\sim 0.5$.      
   
\section{\label{conclusion}Conclusion}

The 3-3-1 and 3-3-1-1 models have been well established. As a result of anomaly cancelation, one of quark families transforms differently from the other two. This leads to dangerous FCNCs due to nonuniversal couplings of $Z',Z''$ with quarks, confronting the current collider bounds. This work show that if the 3-3-1-1 unification is at a very high energy, it imprints at TeV a new neutral gauge boson, $Z'$, conserving flavors, avoiding the above issue. The interest is associated with a dark photon $Z'$ obtained by assigning the dark charge $D=-2/\sqrt{3}T_8+G$ to exotic quarks and leptons, whereas the usual fermions possess $D=0$. 

Different from the usual setup of dark physics, the current dark photon $Z'$ imprinted by 3-3-1-1 breaking has chiral couplings with normal fermions proportional to hypercharge, i.e. $\sim c_\theta \bar{f}\ga^\mu Y f Z'_\mu$, without necessity of any kinetic mixing. Studying various bounds yield that the dark photon should have a mass in TeV regime. 

Additionally, the dark photon $Z'$ negligibly contributes to dark matter observables. Instead, the Higgs field associated with it, i.e. the one ($H'$) that breaks dark charge induces dark photon mass, manifestly governs the dark matter relic density and direct detection. Last, but not least, the $H'$ resonance is necessarily to set the correct relic density. Consequently, the dark matter mass is suitably given at TeV regime.           

\section*{Acknowledgements}

This research is funded by Vietnam National Foundation for Science and Technology Development (NAFOSTED) under grant number 103.01-2023.50.

\end{document}